\documentclass[mathleft]{an}
\usepackage{graphicx}
\usepackage{times}
\usepackage{natbib}
\bibpunct{(}{)}{;}{a}{}{,}
\def\aap{A\&A~}
\def\physrep{Phys.~Rep.~}

\begin{document}

\Pagespan{789}{}
\Yearpublication{2006}%
\Yearsubmission{2005}%
\Month{11}%
\Volume{999}%
\Issue{88}%

\title{Merger Shocks in Abell 3667 and the Cygnus~A Cluster}

\author{Craig L. Sarazin\inst{1}\fnmsep\thanks{Corresponding author:
  \email{sarazin@virginia.edu}\newline}
\and
Alexis Finoguenov\inst{2,3}
\and
Daniel R. Wik\inst{4}%
}
\titlerunning{Merger Shocks in A3667 and Cygnus~A}
\authorrunning{Sarazin, Finoguenov, \& Wik}
\institute{Department of Astronomy,
University of Virginia,
P.O. Box 400325,
Charlottesville, VA 22904-4325, U.S.A
\and
Department of Physics, University of Helsinki, Gustaf H\"allstr\"omin katu 2a,
FI-00014 Helsinki, Finland
\and
Center for Space Science Technology, University of Maryland Baltimore
County, 1000 Hilltop Circle, Baltimore, MD 21250, USA
\and
Astrophysics Science Division, 
NASA/Goddard Space Flight Center,
Greenbelt, MD 20771, USA
}

\received{30 May 2005}
\accepted{11 Nov 2005}
\publonline{later}


\abstract{%
We present new XMM-Newton observations of the northwest (NW) radio relic region in the cluster
Abell 3667.
We detect a jump in the X-ray surface brightness and X-ray temperature at the sharp outer
edge of the radio relic which indicate that this is the location of a merger shock with
a Mach number of about 2.
Comparing the radio emission to the shock
properties implies that approximately 0.2\% of the dissipated shock kinetic
energy goes into accelerating relativistic electrons.
This is an order of
magnitude smaller than the efficiency of shock acceleration in many Galactic
supernova remnants, which may be due to the lower Mach numbers of cluster
merger shocks.
The X-ray and radio properties indicate that the magnetic field
strength in the radio relic is $\ga$3 $\mu$G, which is a very large field at a
projected distance of $\sim$2.2 Mpc from the center of a cluster.
The radio
spectrum is relatively flat at the shock, and steepens dramatically with
distance behind the shock. This is consistent with radiative losses by the
electrons and the post-shock speed determined from the X-ray properties.
The Cygnus~A radio source is located in a merging cluster of galaxies.
This appears to be an early--stage merger.
Our recent Suzaku observation confirm the presence of a hot region between the two subclusters which
agrees with the predicted shocked region.
The high spectral resolution of the CCDs on Suzaku allowed us to measure the radial component of 
the merger velocity, $\Delta v_r \approx 2650$ km/s.%
}

\maketitle

\section{Introduction} \label{sec:intro}
Many clusters of galaxies appear to
be forming at the present time through massive cluster mergers.
Merger shocks driven into the intracluster gas are the primary heating
mechanism of the gas in massive clusters.
Chandra and XMM-Newton
X-ray observations have provided beautiful images and spectra of merger
hydrodynamical effects, including cold fronts and merger bow shocks
\citep[e.g.,][]{MV07}.
However, the number of clusters with well-observed merger shocks is
limited.
Here, we present XMM-Newton and Suzaku observations of merger
shocks in Abell~3667 and the cluster hosting the bright radio source Cygnus~A.

\section{XMM-Newton Observations of a Shock at the NW Radio Relic in Abell~3667}
Abell~3667 is an X-ray bright, low redshift ($z = 0.0552$) cluster which is undergoing a violent
merger
\citep*{MSV99, VMM01b}.
The cluster contains a spectacular pair of curved cluster radio relics \citep{RWH+97} at large distances
from the cluster center.
The NW radio relic is the brightest diffuse cluster radio source
(e.g., cluster relic or halo) in the sky with a flux of 3.7 Jy at 20 cm
\citep{Joh-Hol04}.
It is located at a projected distance of $\sim$2.2 Mpc from the cluster center.
Previous Chandra and XMM-Newton observations of Abell~3667 had concentrated on the central regions
of the cluster where there is a dramatic cold front.
Thus, in Cycle 7 we obtained a 55 ksec XMM-Newton observation of the NW Relic region
\citep{FSN+10}.
More recently, in Cycle 9 we were approved for a 311 ksec XMM-Newton Large Project which gave
a deeper image of the NW Relic, as well as the cluster around it.

\begin{figure}[t]
\begin{center}
\includegraphics[width=0.8\linewidth]{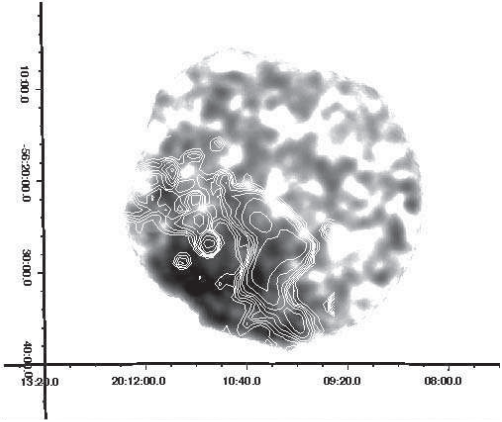}
\end{center}
\vskip0.1\baselineskip
\caption{XMM-Newton image (pn+MOS) from our Cycle 7 observation of the region around the NW radio
  relic in A3667 in the 1.6-4 keV energy band
  \citep[darker $=$ brighter;][]{FSN+10}.
  The X-ray point sources have
  been removed, and the image was smoothed with a 32\arcsec\ Gaussian.  The
  white contours show the radio emission from the relic.}
\label{fig:Sarazin_a3667_xmm}
\end{figure}

Our XMM-Newton data shows a sharp surface brightness discontinuity at the outer edge of the NW Relic
(Fig.~\ref{fig:Sarazin_a3667_xmm}), implying a jump in the gas density.
Spectral observations show a corresponding increase in the gas temperature, indicating that this is a shock.
From the X-ray surface brightness on either side of the surface brightness discontinuity, we estimate that
the shock compression is
$C \equiv n_{e,2} / n_{e,1} = 1.94 \pm 0.19$, where the subscripts 1 and 2 refer to pre-shock and post-shock gas.
For a $\gamma = 5/3$ shock, the shock jump conditions give
\begin{equation}
\frac{1}{C} = \frac{3}{4 {\cal{M}}^2} + \frac{1}{4} \, .
\label{eq:shock_c}
\end{equation}
This gives a shock Mach number of ${\cal{M}} = 1.68 \pm 0.16$.
The effects
of projection, the finite width of the regions used to determine the surface
brightness, the XMM PSF, and the likely curvature of the shock front mean
that this is probably an underestimate.

One can also estimate the Mach number from the observed temperature jump at the shock
front,  which is  $ T_2 / T_1 = 2.68 \pm 1.19$. 
For a
$\gamma = 5/3$ shock, the jump conditions give
\begin{equation}
\frac{T_2}{T_1} = \frac{5 {\cal{M}}^4 + 14 {\cal{M}}^2 - 3}{16 {\cal{M}}^2} \, ,
\label{eq:shock_t}
\end{equation}
which gives ${\cal{M}} = 2.43 \pm 0.77$.
Because of the larger errors in the temperature, the uncertainty in this value is large.

If one takes the weighted mean of the Mach number from the shock compression
and shock temperature front, one finds $ {\cal{M}} = 1.71 \pm 0.16$.  Using
this value, the shock compression is $C = 1.97 \pm 0.19$.  The pre-shock
sound speed is $c_{s1} = 710 \pm 110$ km s$^{-1}$ which gives a shock speed
of $v_s = {\cal{M}} c_{s1} = 1210 \pm 220$ km s$^{-1}$.

Could this merger shock be accelerating or re-accelerating the relativistic
electrons in the NW radio relic?
We start by estimating the energy dissipated in the shock.
As a simple estimate of this, we take the change
in the kinetic energy flux across the shock.
This is given by
\begin{equation}
\Delta F_{\rm KE} = \frac{1}{2} \rho_1 v_s^3 \left( 1 - \frac{1}{C^2} \right) \, ,
\label{eq:shock_e}
\end{equation}
where $\rho_1$ is the pre-shock mass density in the gas.
Using our derived
values for the shock properties gives $\Delta F_{\rm KE} \approx 9 \times
10^{-5} \, {\rm erg} \, {\rm cm}^{-2} \, {\rm s}^{-1}$.  The width of the
relic from northeast to southwest is roughly 26\farcm3 or 1.63 Mpc.  Taking
the area of the shock perpendicular to the flow as a circle with this
diameter gives a perpendicular area of 2.09 Mpc$^2$.  With this size, the
total rate of conversion of shock kinetic energy is
\begin{equation}
\frac{d E_{\rm KE}}{dt} \approx 1.8 \times 10^{45} \, {\rm erg} \, {\rm s}^{-1} \, .
\label{eq:shock_dedt}
\end{equation}

We now compare the rate of shock energy dissipation to the energy required to
power the NW Relic.  Given the steep radio spectrum of the relic, it
appears that the relativistic electrons are losing energy due to radiation.
Thus, in steady-state the power needed to accelerate the electrons is equal to
the total luminosity of the relic.
Integrating up the observed radio spectrum and then adding in inverse Compton (IC) losses
\citep[see][]{FSN+10}
gives a total nonthermal luminosity of
$L_{\rm NT} \approx 3.8 \times 10^{42} [ ( 3.6 \, \mu{\rm G} / B )^2 + 1 ]$ erg s$^{-1}$,
where $B$ is the magnetic field.
If this energy is provided by the
shock acceleration of electrons, then the efficiency
is
\begin{equation}
\epsilon \equiv \frac{\frac{d E_{e}}{dt}}{\frac{d E_{\rm KE}}{dt}}
\approx \frac{L_{\rm NT}}{\frac{d E_{\rm KE}}{dt}}
\approx 0.0021 \left[ \left( \frac{3.6 \, \mu{\rm G}}{B} \right)^2 + 1 \right]
\,  .
\label{eq:shock_eff}
\end{equation}
In \citet{FSN+10}, we show that the magnetic field in the relic is at least $ B \ga 3$
$\mu$G, so that the correction factor in equation~(\ref{eq:shock_eff}) is
between one and 2.5.
Note that this acceleration efficiency is about one order of magnitude
smaller than the values of a few percent usually inferred from the radio
emission by Galactic supernova remnants.
This might be due to the relatively low Mach number of this shock compared to those in
supernova remnants.

\begin{figure}[h]
\begin{center}
\includegraphics[width=0.8\linewidth]{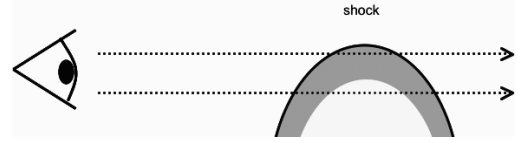}
\end{center}
\caption{Effect of three-dimensional shape of merger shock on the variation
of the radio surface brightness and spectrum with projected post-shock distance.
Since our line of sight intersects the front and back shock surfaces even at a large
projected post-shock distance, the surface brightness doesn't drop or spectrum steepen as
much as expected for a two-dimensional shock model with radiative losses.}
\label{fig:Sarazin_shock_3d}
\end{figure}

The NW Relic has a very sharp outer edge, fades away to
the southwest rather than ending abruptly, and 
the radio spectrum steepens with projected distance from the
outer edge of the relic \citep{RWH+97,Joh-Hol04}.
These are all expected for shock acceleration.
The change in the spectrum would be primarily due to 
the radiative losses as the electrons are advected away from the shock.
If the radio spectrum steepens at a frequency
$\nu_b$, the radiative age for electrons which produce the break is
\citep{vdLP69}
\begin{eqnarray}
t_{\rm rad} & \approx & 1.3 \times 10^8
\left( \frac{\nu_b}{1.4 \, {\rm GHz}} \right)^{-1/2}
\left( \frac{B}{3 \, \mu{\rm G}} \right)^{-3/2} \nonumber \\
& & \qquad \times \left[ \left( \frac{3.6 \, \mu{\rm G}}{B} \right)^2 + 1 \right]^{-1}
\, {\rm yr}
\, .
\label{eq:trad}
\end{eqnarray}
The speed of the post-shock material relative to the shock is $v_2 = v_s / C
\approx 610$ km s$^{-1}$.  
At the redshift of the cluster, the electrons will have moved a projected distance of
\begin{eqnarray}
\theta_{\rm rad} & \approx & 1\farcm3
\left( \frac{\nu_b}{1.4 \, {\rm GHz}} \right)^{-1/2}
\left( \frac{B}{3 \, \mu{\rm G}} \right)^{-3/2} \nonumber \\
& & \qquad
\times \left[ \left( \frac{3.6 \, \mu{\rm G}}{B} \right)^2 + 1 \right]^{-1}
\cos \phi
\,
\label{eq:thetarad}
\end{eqnarray}
where $\phi$ is the angle between the central shock normal and the plane of
the sky. This is roughly consistent with the observed thickness of the layer
at the front outer edge of the radio relic where the spectrum between 20 and
13 cm is observed to steepen dramatically \citep{RWH+97}.

On the other hand, the radio spectrum only steepens by $\sim$1, and the relic remains
visible for a total width of $\sim$8\arcmin\ at 1.4 GHz.
This is expected since the shock is a curved three-dimensional structure,
and far from the NW edge of the relic, we are still seeing radio
emission from relativistic electrons which have just been accelerated
and which are located at the front or back edge of the convex shock region
(Fig.~\ref{fig:Sarazin_shock_3d}).

First order Fermi acceleration gives relativistic electrons with a power-law
spectrum $n(E) \, d E \propto E^{-p} \, d E$, where the power-law index is
$p = ( C + 2 ) / ( C - 1)$, $C$ is the shock compression, and the corresponding radio spectral
index near the shock should be $\alpha = - ( p - 1 ) /2$. 
This would give $\alpha \approx -1.55$
at the shock. The observed radio spectrum is flatter ($\alpha \approx -0.7$)
near the outer edge of the relic
\citep{RWH+97,Joh-Hol04}. This might indicate that we have underestimated
the shock compression, or that the simplest shock acceleration model for the
relic does not apply.

\begin{figure}[h]
\vspace{-0.5\baselineskip}
\begin{center}
\includegraphics[width=0.8\linewidth]{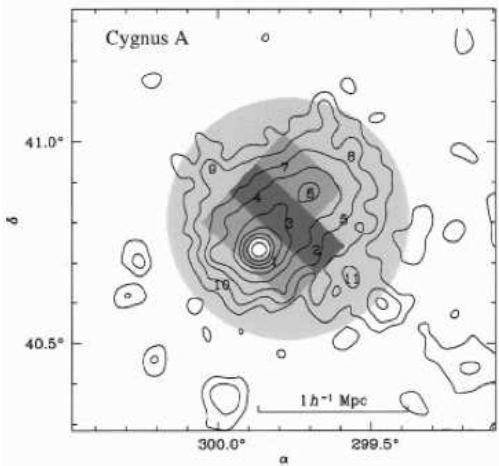}
\end{center}
\caption{Grey-scale is the ASCA X-ray temperature map of the Cygnus~A cluster (darker $=$ hotter). 
The contours are the ROSAT image.
Note the bimodal structure of this merging cluster, and the hot shock regions between the
two subclusters.
Adapted from
\citet*{MSV99}.}
\label{fig:Sarazin_Cyg-A-ASCA}
\end{figure}

The limits on the hard X-ray IC emission from the relic from  our XMM-Newton and
Suzaku observations
\citep{Nak+09,FSN+10}
imply that the magnetic field within the relic is
$B \ge 3$ $\mu$G.
This limit is just consistent with Faraday
rotation measure estimates from the the observation of background radio
galaxies in the zone close to the relic \citep{Joh-Hol04}.
This is a remarkably strong magnetic field at a projected radius of
$\sim$2.2 Mpc from the center of a cluster.
It implies that the nonthermal contribution to the pressure in the relic is
$\ga$20\% of the thermal pressure.
On the other hand, this is the brightest radio relic or halo in the sky,
so this value is unlikely to be typical for the virial regions of clusters.

\section{Suzaku Measurement of the Merger Shock Velocity in
the Cygnus~A Cluster}

Cygnus~A is one of the most luminous radio galaxies in the nearby
universe.
Originally, it was thought to lie in a relatively isolated giant elliptical
galaxy.
However, X-ray observations with the Einstein Observatory established
that Cygnus~A is actually located in a moderately rich cluster
\citep{AFE+84},
and that the radio source and host galaxy are at the center of a
cool core.
Because of its relatively low Galactic latitude, the extensive galaxy
population of the cluster was only discovered more recently
\citep*{LOM05}.

\begin{figure}
\begin{center}
\includegraphics[width=0.8\linewidth]{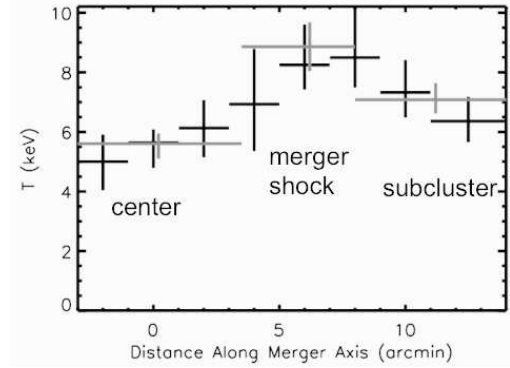}
\end{center}
\caption{Suzaku temperatures in the Center, Subcluster, and Merger Shock regions of the Cygnus~A cluster as a function
of distance along the merger axis.
Note that the shock region is hot.}
\label{fig:Sarazin_Cyg-A_T}
\end{figure}

ASCA and Chandra X-ray images and spectra have shown that the
Cygnus~A cluster is undergoing a merger with a particularly simple geometry
\citep{MSV99,SWA+02}.
The cluster consists of two subclusters, one centered on the Cygnus A
radio galaxy, the other centered on an enhancement in the X-ray
brightness to the northwest of the radio galaxy
(Fig.~\ref{fig:Sarazin_Cyg-A-ASCA}).
If one excludes the flux from the cooling core, the two subclusters
appear to be similar in their X-ray properties.
The ASCA temperature map
\citep{MSV99}
and Chandra image
\citep{SWA+02}
show a region of increased temperature and entropy located between the two
subcluster centers and oriented perpendicular to the elongation of the
cluster
(Fig.~\ref{fig:Sarazin_Cyg-A-ASCA}),
which is likely to be a merger shock.
This suggests that this is a fairly symmetric, straight-on merger
at an early stage where the merger shocks have not yet passed the subcluster
centers.
%
The measured radial velocity distribution of the galaxies shows a
strongly bimodal distribution, with 
$1600 \la \Delta v_r \la 2600$ km/s
\citep{LOM05}.

In Cycle~3, we obtained a Suzaku observation of 44 ksec of the Cygnus~A cluster.
The pointing was selected to include both subclusters and the merger shock region.
The primary goals were to use the excellent CCD spectral resolution of Suzaku to directly
measure the merger and shock velocities using the Doppler shifts of X-ray lines, and to confirm
the temperature structure of the shock region.
We have analyzed the Suzaku data along with an existing XMM-Newton observation from
the archive.
For details of the analysis and results, see
\citet{WS12}.

\begin{figure}
\begin{center}
\includegraphics[width=0.8\linewidth]{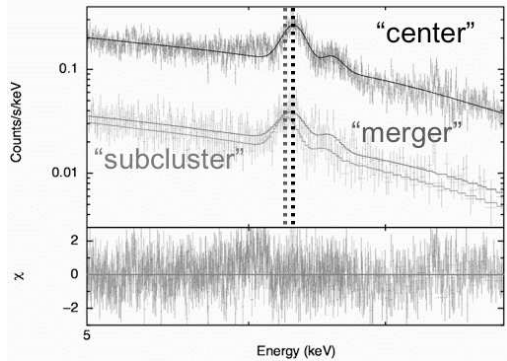}
\end{center}
\caption{Suzaku spectra of the Cygnus~A cluster near the Fe-K line.
There is a clear shift downward in the line energy in going from the Center to Merger Shock to Subcluster regions.}
\label{fig:Sarazin_Cyg-A_spectra}
\end{figure}

\begin{figure}[b]
\begin{center}
\includegraphics[width=0.8\linewidth]{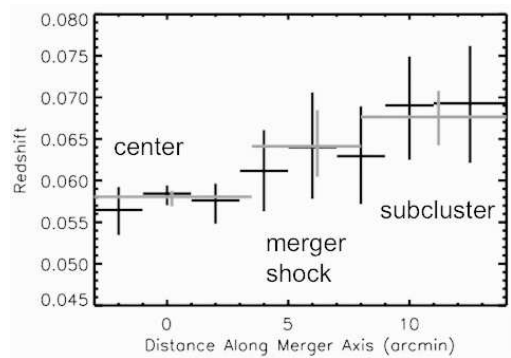}
\end{center}
\caption{Suzaku redshifts in the Cygnus~A cluster.
Note the increase in redshift with distance along the merger axis in going from the
Center to Merger Shock to Subcluster regions.}
\label{fig:Sarazin_Cyg-A_z}
\end{figure}

The Suzaku spectra were accumulated in three main regions:
the ``center'' region included the subcluster containing the radio source, but excluding the region immediate
surrounding the radio source and cool core;
the ``subcluster'' region containing the second merging subcluster to the northwest;
and the ``merger shock'' region containing the likely merger shock, based on the 
ASCA and Chandra X-ray images.
Each of these regions was subdivided in 3 regions.
The results are summarized in Figures~\ref{fig:Sarazin_Cyg-A_T}, \ref{fig:Sarazin_Cyg-A_spectra}, \&
\ref{fig:Sarazin_Cyg-A_z}.

The spectra confirm that the shock region is hotter than either subcluster, as expected if it is shocked
gas between the two subclusters
(Fig.~\ref{fig:Sarazin_Cyg-A_T}).
The temperature of the Center region is probably underestimated as a result of the difficulty in
excising the cool core due to the large Suzaku PSF.

The spectra in the Fe K line region
(Figs.~\ref{fig:Sarazin_Cyg-A_spectra} \& \ref{fig:Sarazin_Cyg-A_z})
show that the redshift of the X-ray gas increases very substantially in going from the Center to the
Merger Shock to the Subcluster region.
The sense of the difference implies that the Subcluster is falling in toward the Center from the foreground.
The radial component of the merger velocity (the difference between the two subclusters) is
$\Delta v_r = (2650 \pm 900 )$ km/s (90\%), which is consistent with the bimodal galaxy velocity difference.
Applying the Rankine-Hugoniot shock jump conditions to the de-projected temperature increase
in the Merger Shock region,
we estimate a merger velocity of 2400--3000 km/s.
Combining with our $\Delta v_r$ yields an angle $<$54$^\circ$ between our line-of-sight and the merger axis.
On the other hand, the bimodal X-ray distribution implies that this angle is not small (perhaps
$\sim$45$^\circ$).

\section{Conclusions}

We have presented XMM-Newton observations of a merger shock in Abell~3667.
Most of the properties of this shock and the NW radio relic are consistent with the relic being
due to particle (re)acceleration by the shock.
The observations imply a very strong magnetic field ($B \ge 3$ $\mu$G)
at a large projected radius ($\sim$2.2 Mpc, approximately the virial radius) from the center of the cluster.

In the Cygnus~A cluster, we have made one of the first direct detections of a merger velocity using
the Doppler shifts of X-ray lines from the intracluster gas.
The observed radial component of the merger velocity is quite large,
$\Delta v \approx 2650$ km/s.

\acknowledgements
CLS thanks the Institute for Astro- and Particle Physics at the University of Innsbruck and the
Eramus Mundus Program of the European Commission for their hospitality and
support.
This work was primarily funded by NASA ADAP Grant NNX11AD15G
and NASA Suzaku Grant NNX09AH25G, but also by Chandra Grant
GO1-12169X.




\end{document}